\documentclass[12pt,a4paper]{article}

\usepackage{epsfig}
\usepackage{amstex}

\def\nostrocostrutto#1\over#2{\mathrel{\mathop{\kern 0pt \rlap 
  {\raise.2ex\hbox{$#1$}}}
  \lower.9ex\hbox{\kern-.190em $#2$}}}

\newcommand{\eref}[1]{(\ref{#1})}      

\newcommand{\Nq}{{\mathcal N}_q}
\newcommand{\Ng}{{\mathcal N}_g}
\newcommand{\Ord}{{\mathcal O}}

\newcommand{\MeV}{{\rm MeV}}

\newcommand{\Lpm}{\Lambda}

\catcode`@=11
\newcount\@tempcntc
\def\@citex[#1]#2{\if@filesw\immediate\write\@auxout{\string\citation{#2}}\fi
  \@tempcnta\z@\@tempcntb\m@ne\def\@citea{}\@cite{\@for\@citeb:=#2\do
    {\@ifundefined
       {b@\@citeb}{\@citeo\@tempcntb\m@ne\@citea\def\@citea{,}{\bf ?}\@warning
       {Citation `\@citeb' on page \thepage \space undefined}}%
    {\setbox\z@\hbox{\global\@tempcntc0\csname b@\@citeb\endcsname\relax}%
     \ifnum\@tempcntc=\z@ \@citeo\@tempcntb\m@ne
       \@citea\def\@citea{,}\hbox{\csname b@\@citeb\endcsname}%
     \else
      \advance\@tempcntb\@ne
      \ifnum\@tempcntb=\@tempcntc
      \else\advance\@tempcntb\m@ne\@citeo
      \@tempcnta\@tempcntc\@tempcntb\@tempcntc\fi\fi}}\@citeo}{#1}}
\def\@citeo{\ifnum\@tempcnta>\@tempcntb\else\@citea\def\@citea{,}%
  \ifnum\@tempcnta=\@tempcntb\the\@tempcnta\else
   {\advance\@tempcnta\@ne\ifnum\@tempcnta=\@tempcntb \else \def\@citea{--}\fi
    \advance\@tempcnta\m@ne\the\@tempcnta\@citea\the\@tempcntb}\fi\fi}
\catcode`@=12

\begin{document}

\setcounter{page}{0}
\thispagestyle{empty}

\title{Perturbative QCD description of multiparticle correlations 
in quark and gluon jets}

\author{Sergio Lupia\thanks{E-mail address: lupia@mppmu.mpg.de} }

\date{{\normalsize\it Max-Planck-Institut 
f\"ur Physik (Werner-Heisenberg-Institut), \\
F\"ohringer Ring 6, 80805 M\"unchen, Germany}}

\maketitle
\thispagestyle{empty}

\begin{abstract}
The QCD evolution equations in Modified Leading Log Approximation 
for the factorial moments of the
multiplicity distribution in quark and gluon jets 
are numerically solved with initial conditions at threshold  
by fully taking into account the energy conservation law. 
After applying Local Parton Hadron Duality as
hadronization prescription, a consistent quantitative description of
available experimental data 
for factorial cumulants and factorial moments of arbitrary order and for 
their ratio both in quark and gluon jets and in $e^+e^-$ annihilation 
is achieved. 
\end{abstract}

\vspace{-15cm}

\rightline{MPI-PhT/98-49}
\rightline{June 26th, 1998} 

\vspace{19cm} 

\noindent 
Keywords: Perturbative QCD, jets, multiparticle dynamics, 
hadronization, $e^+e^-$ annihilation, multiplicity moments. \\ 
\noindent 
PACS Numbers: 12.38.Bx, 13.65.+i. 

\newpage

\section{Introduction}

The analytical perturbative approach to multiparticle production 
in jets \cite{dkmt} is widely used to describe
 inclusive observables  in $e^+e^-$ annihilation. 
In this framework, a parton cascade is evolved down to small scales
of a few hundred MeV for the transverse momentum cutoff $Q_0$, 
and partonic predictions are directly compared to hadronic observales, 
according to the notion of local parton hadron duality (LPHD) \cite{lphd}.
The description of the parton cascade is given within the 
so-called modified leading logarithmic approximation (MLLA),  
which takes into account the leading double logarithmic and next to leading
single logarithmic terms; in this approximation,  
QCD coherence is included by means of angular ordering, and the running of
the coupling $\alpha_s$ at one-loop level 
and energy conservation effects are included as well. 

This approach has been applied to a large variety of
experimental data and found to be rather successful 
(see \cite{ko} for a recent review). In particular, 
by considering predictions which fully take 
into account the effects of the energy conservation law, 
a quantitative description of available experimental data  for
single particle observables, like for instance the momentum spectrum
\cite{lo} and both the jet and the particle multiplicity~\cite{lomult},  
has been achieved. 
However, in the case of observables related to genuine multiparticle correlations, 
like factorial cumulants and factorial moments of the multiplicity
distributions, the available theoretical predictions~\cite{dremin}, which 
include only partially the energy conservation effects, 
do not quantitatively  reproduce the experimental data\cite{opalmom}.  

In this letter we extend the results of ~\cite{lomult} for the average
multiplicity and we present a numerical solution of the complete MLLA evolution
equation of QCD for factorial moments of any order with the full inclusion
of energy conservation effects. In this way, we can  
quantitatively describe within the purely perturbative approach 
recent experimental data on multiparticle correlations both in single quark and
gluon jets and in $e^+e^-$ annihilation, thus lending further support to the
analytical perturbative approach to multiparticle production.

\section{The theoretical framework}

In order to study genuine multiparticle correlations in a quark or gluon jet, 
it is convenient to study the properties of the (unnormalized) 
factorial moments of order $q$, given in terms of the multiplicity 
distribution $P_n$ by 
\begin{equation} 
\tilde F^{(q)}=  \sum_{n=q}^{\infty} n(n-1)\dots(n-q+1) P_n  ,
\label{facmom}
\end{equation} 
and the (unnormalized) factorial cumulants of order $q$, related to the factorial moments 
by a cluster expansion: 
\begin{equation}
\tilde  K^{(q)} = \tilde F^{(q)} - \sum_{i=1}^{q-1} {q-1 \choose i} 
\tilde K^{(q-i)} \tilde F^{(i)} .  
\label{faccum}
\end{equation}
Factorial moments and factorial cumulants are integrals of $q$-particle
inclusive distributions and $q$-particle correlations functions,
respectively. 
Normalized observables are obtained by using the average multiplicity 
$\bar n = F^{(1)} = K^{(1)}$: 
\begin{equation}
F^{(q)} = \frac{\tilde F^{(q)}}{[\bar n]^q} \quad , \quad 
K^{(q)} = \frac{\tilde K^{(q)}}{[\bar n]^q}
\end{equation}  
Also the ratio  of factorial cumulants over
factorial moments
\begin{equation}
  H_q = K_q / F_q                                  \label{hmom}
\end{equation} 
has been widely studied.

One usually also considers the generating functions 
of the multiparton final states in quark and gluon jets, $Z_a(u)$, 
from which the factorial moments can be obtained via a Taylor expansion 
around $u=1$ (here $a = q,g$ for quark and gluon jet respectively):
\begin{equation}
\tilde F_a^{(q)} = \frac{d^q Z_a(u)}{du^q} \mid_{u=1},
\label{den}       
\end{equation}    
and similarly for the cumulant moments 
\begin{equation}  
\tilde K_a^{(q)} = \frac{d^q \ln Z_a(u)}{du^q} \mid_{u=1} .      
\label{con}       
\end{equation}    

The evolution equations for the generating functions $Z_a(u)$ 
have been derived  within a probabilistic description
 of the parton splitting processes $A\to B + C$, with the inclusion 
of angular ordering, energy conservation 
and the running coupling at the one-loop order\cite{dkmt,dok,do} 
(we explicitly show the energy dependence and 
drop the label $u$ everywhere for clarity): 
\begin{eqnarray}
\left\{ \begin{array}{l} 
\displaystyle\frac{dZ_g (\eta)}{d\eta} = \displaystyle\int_{z_c}^{1-z_c} dz 
      \displaystyle\frac{\alpha_s(\tilde k_\perp)}{2\pi}[\Phi_{gg}^{asy}(z)
      \{Z_g(\eta+\ln z) Z_g(\eta+\ln (1-z))-Z_g(\eta)\}  + \\
   \qquad \qquad  n_f \Phi_{gq}(z)
       \{Z_q(\eta+\ln z) Z_q(\eta+\ln (1-z))-Z_g(\eta)\}] \label{eveq} 
\vspace{0.3cm}   \\ 
\displaystyle\frac{dZ_q (\eta)}{d\eta} = \displaystyle\int_{z_c}^{1-z_c} dz
      \displaystyle\frac{\alpha_s(\tilde k_\perp)}{2\pi}\Phi_{qg}(z)
      \{Z_g(\eta+\ln z) Z_q(\eta+\ln (1-z))- Z_q(\eta)\} 
\end{array} \right.
\end{eqnarray}
Here the evolution variable $\eta=\ln \frac{\kappa}{Q_0}$ 
is related to the jet virtuality $\kappa=Q\sin \frac{\Theta}{2}$,           
where $Q = 2 E$ is the hard scale involved in the process, and $E$ is 
in this case the jet energy,       
$\Theta$ denotes the maximum  angle between the outgoing partons $B$
and $C$ and $Q_0$ is the infrared cutoff in transverse momentum.

The splitting functions $\Phi_{AB}$ for parton splittings 
$A\to B$~\cite{dglap} are taken with normalization as in
\cite{dkmt,lomult}; here $\Phi_{gg}^{asy}(z)=(1-z)\Phi_{gg}(z)$ 
can replace $\frac{1}{2}\Phi_{gg}(z)$ thanks to symmetry properties 
of the kernel~\cite{do}.
 $N_C$ and $n_f$ denote the number of colours and flavours. 
The coupling is given by 
$\alpha_s^{(n_f)}(\tilde k_\perp)=2\pi/(b \ln (\tilde k_\perp/\Lpm))$
with $b=(11N_C-2n_f)/3$.  
$\alpha_s(\tilde k_\perp)^{-1}$ evolves  
with $\tilde k_\perp$ with a smooth treatment of 
the heavy quark thresholds at the effective mass 
 $m_i^* = e^{5/6} m_i/2 \simeq 1.15 m_i$ as in~\cite{lomult}. 
The boundaries of the integrals are determined by the lower cutoff 
in the transverse momentum measure $\tilde k_\perp$ 
defined according to  the Durham jet finder algorithm\cite{durham} 
$\tilde k_\perp  =  {\rm min}(z,1-z) \kappa\;\geq\; Q_c$:  
\begin{equation}
z_c  = \frac{Q_c\sqrt{2}}{Q} =\sqrt{2 y_c} = e^{-\eta} . 
 \label{zc}
\end{equation}  

By taking the derivatives now in eq.~\eref{eveq} according to eq.~\eref{den}, 
one then obtains a system of coupled
evolution equations for the unnormalized factorial moments for a single
quark or gluon jet: 
\begin{eqnarray}
\left\{ 
\begin{array}{l} 
\displaystyle\frac{d\Ng (\eta)}{d\eta} = \int_{z_c}^{1-z_c} dz 
      \frac{\alpha_s(\tilde k_\perp)}{2\pi}[\Phi_{gg}^{asy}(z)
      \{\Ng(\eta+\ln z)+\Ng(\eta+\ln (1-z))-\Ng(\eta)\} + \\
 \hspace{2.0cm}  n_f \Phi_{gq}(z)
       \{\Nq(\eta+\ln z)+\Nq(\eta+\ln (1-z))-\Ng(\eta)\}]  \vspace{0.3cm}
  \\ 
\displaystyle\frac{d\Nq (\eta)}{d\eta} = \int_{z_c}^{1-z_c} dz
      \frac{\alpha_s(\tilde k_\perp)}{2\pi}\Phi_{qg}(z)
      \{\Ng(\eta+\ln z)+\Nq(\eta+\ln (1-z))-\Nq(\eta)\} \vspace{0.3cm} \\ 
 \hspace{2.0cm}     \vdots \vspace{0.3cm}\\ 
\displaystyle\frac{d\tilde F^{(q)}_g (\eta)}{d\eta} = \int_{z_c}^{1-z_c} dz
      \frac{\alpha_s(\tilde k_\perp)}{2\pi}[\Phi_{gg}^{asy}(z) \times \\ 
\hspace{2.0cm}   \{ \displaystyle\sum_{m=0}^q {q \choose m} \tilde F^{(m)}_g(\eta+\ln z) 
\tilde F^{(q-m)}_g(\eta+\ln (1-z))  - \tilde F^{(q)}_g(\eta)\} + \\
\hspace{2.0cm}   n_f \Phi_{gq}(z) \{ \displaystyle\sum_{m=0}^q {q \choose m} \tilde F^{(m)}_q(\eta+\ln z) 
\tilde F^{(q-m)}_q(\eta+\ln (1-z))  - \tilde F^{(q)}_g(\eta)\}] 
 \vspace{0.3cm} \\                                                
\displaystyle\frac{d\tilde F^{(q)}_q (\eta)}{d\eta} = \int_{z_c}^{1-z_c} dz
      \frac{\alpha_s(\tilde k_\perp)}{2\pi}\Phi_{qg}(z) \times \\ 
\hspace{2.0cm}     \{ \displaystyle\sum_{m=0}^q {q \choose m} \tilde F^{(q-m)}_g(\eta+\ln z) 
\tilde F^{(q-m)}_q(\eta+\ln (1-z))- \tilde F^{(q)}_q(\eta)\} 
\end{array} \right.
\label{eveqmom}
\end{eqnarray} 
Since $z_c = e^{-\eta} \leq \frac{1}{2}$, 
the system of differential equations~(\ref{eveqmom}) is 
defined for $\eta\geq \ln 2$ only. 
Its initial conditions  for $0 \le \eta \le \ln 2$  read 
\begin{eqnarray}
\Ng = \Nq &=& 1 \nonumber \\ 
\tilde F^{(q)}_g = \tilde F^{(q)}_q &=& 0 \qquad \mathrm{for} q > 1 
\label{init}
\end{eqnarray}
No analytical solution of the complete equations~(\ref{eveqmom}) with the boundary
conditions~(\ref{init}) has been so far obtained. 
A numerical solution of the two coupled equations for the average
multiplicity has been given in \cite{lomult}. 
For high order factorial moments, both solutions within high energy
approximations at leading\cite{bg} and next-to-leading
order\cite{mw} and a solution of yet higher order 
with partially takes into account 
the energy conservation law\cite{dremin} have been obtained. 
Notice that these solutions do not fulfill the absolute
normalization at threshold, where they reach unphysical negative values for 
the moments. 

Here we numerically solve the system of equations ~\eref{eveqmom} with 
boundary conditions at threshold~\eref{init}  for the first 20
factorial moments. 

\subsection{From one to two hemispheres}

We have so far considered particle production inside 
a single quark or gluon jet. In this framework, a single quark jet is 
defined in an inclusive sense, i.e., it experimentally corresponds  to 
a single hemisphere in a $e^+e^-$ annihilation event. 
At low $cms$ energies this factorization of the two hemispheres does not hold
anymore and nonlogarithmic corrections become important. 
At large $cms$ energies, however, the behaviour of a whole $e^+e^-$
annihilation event is controlled by the generating function
\begin{equation} 
Z_{two-hem}(u) = [Z_q(u) ]^2  \quad \Leftrightarrow \quad 
\ln Z_{two-hem}(u) = 2 \ln [Z_q(u) ]
\end{equation} 
It is then easy to relate the moments for the whole event to the moments 
for one hemisphere that one can calculate according to the evolution
equation~\eref{eveqmom}. One obtains indeed for the factorial cumulants: 
\begin{equation}                                                  
\bar n_{two-hem} = 2 \bar n_q \quad , \quad 
\tilde K^{(q)}_{two-hem} = 2 \tilde K^{(q)}_q  
\label{twohemnonorm}
\end{equation} 
and for normalized cumulants
\begin{equation} 
K^{(q)}_{two-hem} = \frac{ K^{(q)}_q}{2^{q-1}} 
\label{twohem}
\end{equation} 
Factorial moments can then be calculated from the factorial cumulants by
inverting eq.~\eref{faccum}. For instance, for the second order factorial 
moment, one gets: 
\begin{eqnarray}
\tilde F^{(2)}_{two-hem} &=& \bigl[ \bar n_{two-hem} \bigr]^2 + 
 2 \bigl[ \tilde F^{(2)}_q - {\bar n_q}^2 \bigr] \nonumber \\ 
 &=& 2 \bigl[ {\bar n_q}^2 + \tilde F^{(2)}_q   \bigr] 
\label{f2twohem}
\end{eqnarray}
The simple treatment of the whole $e^+e^-$ event in terms of a superposition 
of two independent single quark jets have been improved for the average
multiplicity by explicitly including the  $\Ord(\alpha_s)$ matrix element 
for $e^+e^- \to$ 3 partons\cite{bs,cdfw}. In this way, an improved 
description of data in the low $cms$ energy region could be
achieved\cite{lo}. Since we are here mainly interested 
in the high energy regime of the moments, where the correction is negligible, 
we will use in the following the aforementioned simple ansatz.

\section{Phenomenology of high order correlations}

\subsection{Correlations in single quark and gluon jets}

\begin{table}     
 \begin{center}
 \vspace{4mm}
 \begin{tabular}{||c|c|c||c|c||}
\hline 
$q$ & \multicolumn{2}{c||}{$F_q^{(q)}$ Quark jet} 
& \multicolumn{2}{c||}{$F_g^{(q)}$ Gluon jet} \\ 
 \hline 
  & Exp\cite{opalmom} & Theory & Exp\cite{opalmom} & Theory   \\ 
\hline 
 2 & $1.0820\pm 0.0006\pm 0.0046$ & 1.080 & 
$1.023\pm 0.008 \pm 0.011$  & 1.026 \\ 
 3  & $1.275\pm 0.002\pm 0.017$ & 1.265 & 
$1.071\pm 0.026 \pm 0.034$ & 1.078 \\ 
 4 & $1.627\pm 0.005\pm 0.042$ & 1.600 & 
$1.146\pm 0.059\pm 0.074$ & 1.157  \\ 
 5 & $2.274\pm 0.014\pm 0.093$ & 2.168 & 
$1.25\pm 0.11\pm 0.13$ & 1.268 \\ 
 \hline
 \end{tabular}
 \end{center}
\caption{Experimental data for the 
normalized factorial moments of order $q = 2,\dots,5$ in single
quark and gluon jets of 45.6 and 41.8 GeV respectively~\protect\cite{opalmom}
 are compared 
with our theoretical predictions with parameters 
given in~\protect\eref{results}.} 
\label{tableparameter}
\end{table}

Both the evolution equation~\eref{eveq} and eq.~\eref{eveqmom}   
refer to parton production inside a
single quark or gluon jet in a fully inclusive configuration. 
This is a very important point for the phenomenological application of these
predictions. It is indeed not possible to directly compare our
predictions with data extracted in different configurations, like for
instance data on single quark and gluon jets from 3-jet events with the
Mercedes configuration, where topology has been 
shown to play an important role~\cite{alephtopo,eden} .
However, it is possible to study a particular subsample of 3-jet events, 
 with a very energetic gluon-jet practically back-to-back to the 
quark-antiquark pair, where the experimental configuration is comparable to 
the theoretical one\cite{gary}.             
 This analysis has been performed by the OPAL Collaboration both for the
average multiplicity\cite{opalglu} and for higher order factorial 
moments\cite{opalmom} for quark jets of 45.6 GeV and gluon jets of 41.8 GeV. 
Table~\eref{tableparameter} compares the experimental data for the first
four factorial moments inside a single quark and gluon jet with the
theoretical predictions obtained by numerical solution of eq.~\eref{eveqmom}
with the boundary conditions~\eref{init}. 
The free parameters of the theory, i.e., 
the infrared cutoff $Q_0$, the QCD scale 
$\Lambda$ and the normalization parameter $K_{all}$, have been kept equal 
to the values fixed in \cite{lomult} to describe the 
jet and particle multiplicities: 
\begin{equation}
K_{all}=1, \qquad \Lambda=500~ \MeV
\qquad  Q_0 = 0.507~\MeV  \label{results}
\end{equation}
A very good quantitative agreement between data and the predictions of the
complete theory is achieved, contrary to previous
comparisons with approximate predictions\cite{opalmom}.

\begin{table}     
 \begin{center}
 \vspace{4mm}
 \begin{tabular}{||c||c|c|c|c|c|c||}
\hline
$q$ & 1 & 2 & 3 & 4 & 5 & 6  \\ \hline
$F_q^{(q)}$ & 1 &
0.07431 &  0.01922 & 0.00135 & -0.00211 &
-0.00150  \\ 
$F_g^{(q)}$ & 1 &
0.028737 & 0.000330 & -0.000197 & -0.000125 &
-0.000024 \\ \hline 
$q$ & 7 & 8 & 9 & 10 & 11 & 12 \\ \hline 
$F_q^{(q)}$ & -0.00018 & 
 0.00056 & 0.00052 &
0.00003 &  -0.00041 & -0.00043 \\ 
$F_g^{(q)}$  & 0.000033 & 
0.000004 & -0.000005
& -0.000005 & 0.000007 & -0.000009 \\ \hline 
$q$ & 13  &14 & 15 & 16 & 17 & 18  \\ \hline 
$F_q^{(q)}$  & 0.00004 &
0.00060 & 0.00066 & -0.00023 & -0.00155 & 
-0.00169  \\ 
$F_g^{(q)}$ & -0.000015 & -0.000018 &  0.000012 & 0.000001 & 
-0.000024 & 0.000067 \\
 \hline
 \end{tabular}
 \end{center}
\caption{Predictions for the ratios $H_q$ in single
quark and gluon jets of 45.6 and 41.8 GeV 
respectively with parameters
given in~\protect\eref{results}.}
\end{table}

To be more explicit, Fig. 1 compares the data for
factorial cumulants of order $q$ = 2 up to 5 in single quark jets of 45.6
GeV  and single gluon jets of 41.8 GeV
with the predictions computed in this paper and with the
predictions of \cite{dremin}. It can be seen that the
full inclusion of energy conservation is
very important for these observables and that, by including this effect, 
 the perturbative approach can
quantitatively describe experimental data in this configuration.

It is also possible to obtain predictions for high order moments in
single quark and gluon jets. Predictions for the ratio $H_q$ inside single
quark and gluon jets up to order 18 are shown in Table~2.  
It is particularly interesting to notice that oscillations of the $H_q$
moments are up to two orders of magnitude 
larger in quark jets than in gluon jets. 

\subsection{Correlations in $e^+e^-$ annihilation}

Fig.~2 shows the dependence on $Y = \ln (\sqrt{s}/Q_0)$ of the second order
normalized factorial moment $F^{(2)}$. Experimental data\cite{ms} are
compared with the leading\cite{bg} and next-to-leading predictions\cite{mw} 
and with our predictions, where the result for the full event has been
obtained from the prediction for a single quark-jet according to 
eq.~\eref{twohem} in the following way: 
\begin{equation} 
F^{(2)}_{e^+e^-} = 1 + \displaystyle\frac{K_q^{(2)}}{2}
\end{equation} 
Also in this case, our improved description is in quantitative agreement
with the experimental data and reaches an accuracy comparable to the widely
used Monte Carlo generators. 

In order to investigate whether our predictions give a good description of 
 correlations of yet higher order as well, we have also compared them with 
the data for the ratio of factorial cumulants over factorial moments, $H_q$, as
a function of the order $q$. It has been shown in \cite{hq2} that an
important part of the oscillatory behaviour of the $H_q$ moments actually 
comes from the superposition of two different samples of two-jet and
multi-jets events, an effect which is not taken into account
in our simple treatment of correlations between the two hemispheres according
to ~\eref{twohem}. At the moment, we cannot then aim at a description of 
data for the whole $e^+e^-$ annihilation events~\cite{sld}, which requires an improved treatment
of the composition of the two hemispheres, but we should look at  
data for  two-jet events only. 
We have therefore compared in Fig.~3 our predictions with data 
for the $H_q$ moments extracted as in \cite{hq2} from experimental
data\cite{delphi} on multiplicity distributions in three samples of 
 two-jet events, defined according to the JADE jet finder algorithm with
three different values of the jet resolution parameter $y_{cut}$. 
One can easily see that the theory can describe the gross 
features of the data for correlations of arbitrary order as well. 
In this case, a more quantitative comparison should not be pursued, since data 
 have been extracted by using a different jet finder algorithm 
than the one used in the theory.

\section{Conclusions}

We have numerically solved the  QCD evolution equations in MLLA
for factorial moments of any order in quark and gluon jets, by fully taking
into account the effects of energy conservation law.  
By keeping the same parameters previously fixed in order to describe the
mean jet and particle multiplicities, a quantitative
description of existing data on factorial moments, factorial cumulants and
their ratio both in single quark and gluon jets and in the two-jet sample of
$e^+e^-$ annihilation has been achieved. 
The analytical perturbative approach, based on perturbative QCD 
description of parton evolution assuming Local Parton Hadron Duality, is then
shown to be quantitatively 
successful in describing genuine multiparticle correlations 
inside jets. 
 
\section*{Acknowledgements} 

I thank Wolfgang Ochs for many useful discussions and
suggestions and a careful reading of the manuscript.

\newpage

\section*{Figure Captions}

{\bf Fig. 1}: {\bf a} 
Data on the first 4 normalized factorial cumulants 
for single quark-jets of 45.6 GeV 
as measured by the OPAL collaboration\cite{opalmom} (diamonds) 
compared with predictions by Dremin et al.~\cite{dremin} (triangle) 
and with our predictions (square); 
{\bf b}: same as in {\bf a}, but for single gluon jets of 41.8 GeV. 

\medskip\noindent 
{\bf Fig. 2}: 
Second order normalized factorial moment $F^{(2)}$ as a function 
of $Y = \ln (\sqrt{s}/2Q_0$, with $Q_0$ = 0.507 GeV; comparison of
experimental data\cite{ms} with leading order predictions\cite{bg} (dashed
line), next-to-leading order predictions\cite{mw} (dotted line) and 
our prediction (solid line). 

\medskip\noindent                         
{\bf Fig. 3}:                             
Ratio $H_q$ of factorial cumulants over factorial moments as a function of
the order $q$; data for two-jet events as extracted in
\cite{hq2} from experimental data on MD's measured by DELPHI
Collaboration\cite{delphi} with three different values of jet resolution
parameters within the JADE algorithm ($y_c$ = 0.01 (diamond), 0.02
(triangle) and 0.04 (square)) are compared with our predictions (solid line).

 \newpage

%

\begin{figure}[p]
\begin{center} 
\mbox{
\mbox{\epsfig{file=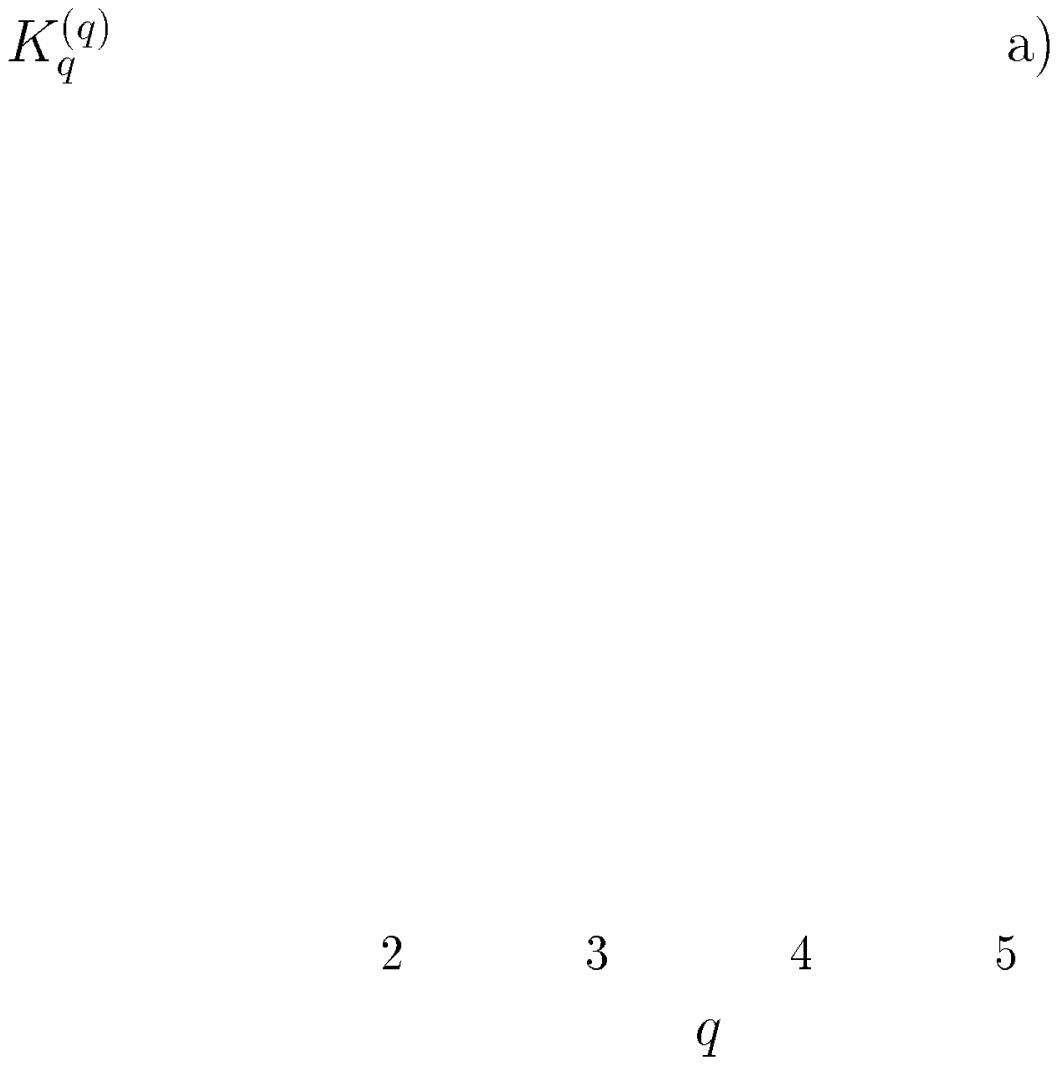,bbllx=4.cm,bblly=1.cm,bburx=4.2cm,bbury=17.cm}}
\mbox{\epsfig{file=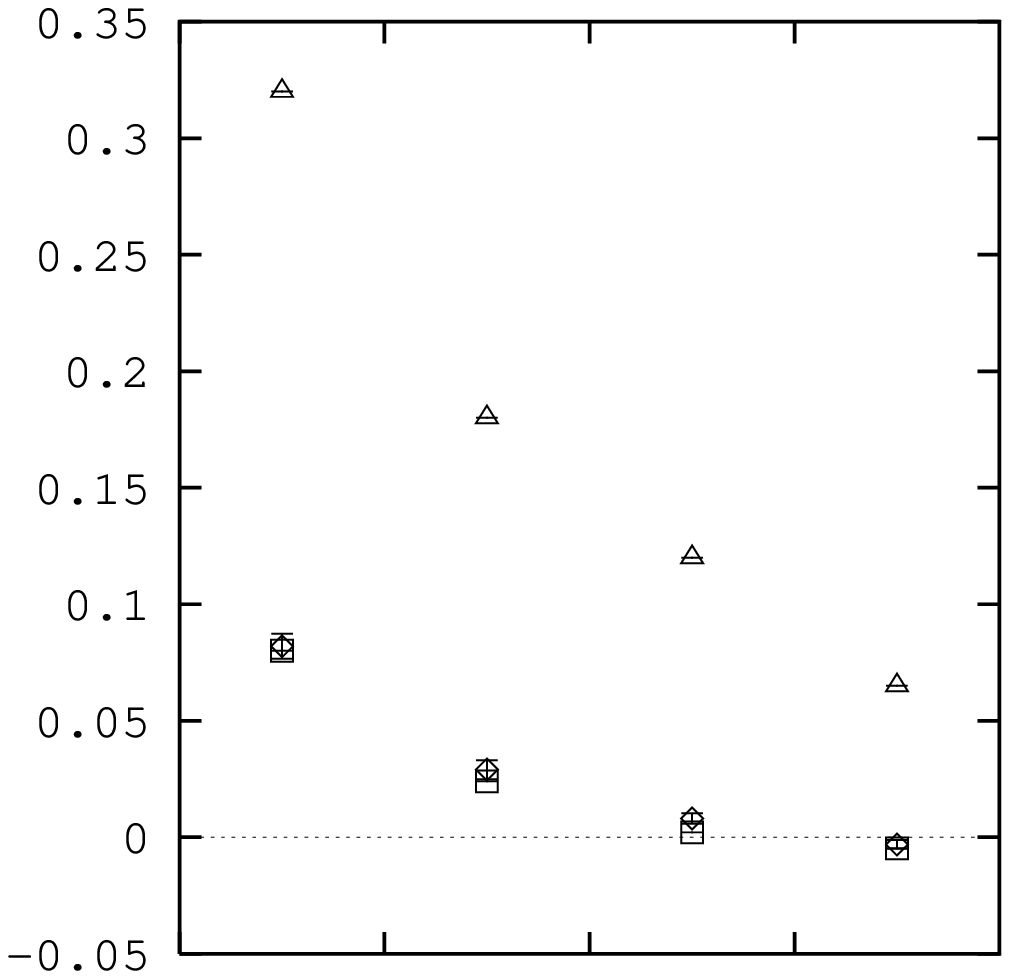,width=19cm,bbllx=2.cm,bblly=1.cm,bburx=21.cm,bbury=17.cm}}
 } \end{center}
\caption{a} 
\label{fig1a}
\end{figure}

\newpage

\setcounter{figure}{0}
\begin{figure}[p]
\begin{center} 
\mbox{
\mbox{\epsfig{file=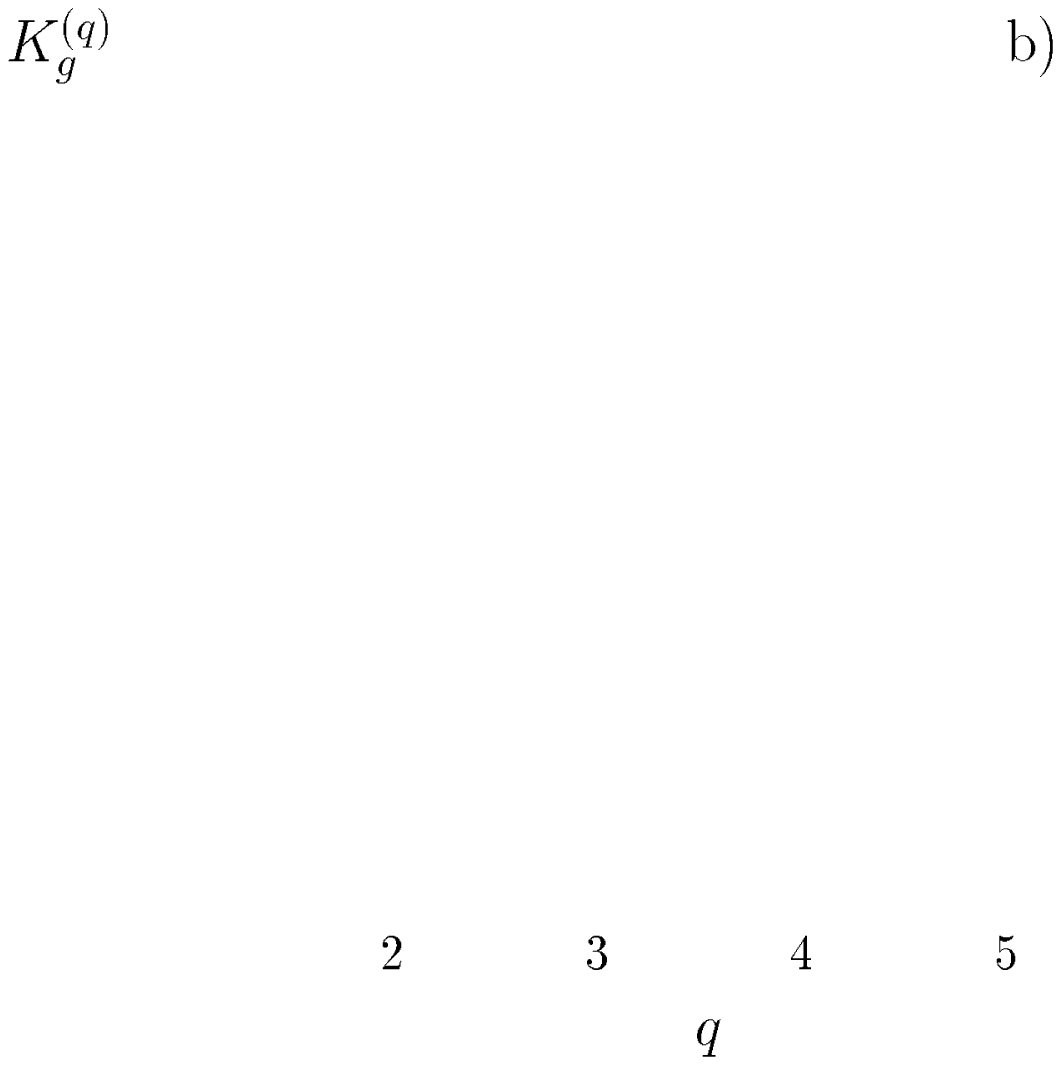,bbllx=4.cm,bblly=1.cm,bburx=4.2cm,bbury=17.cm}}
\mbox{\epsfig{file=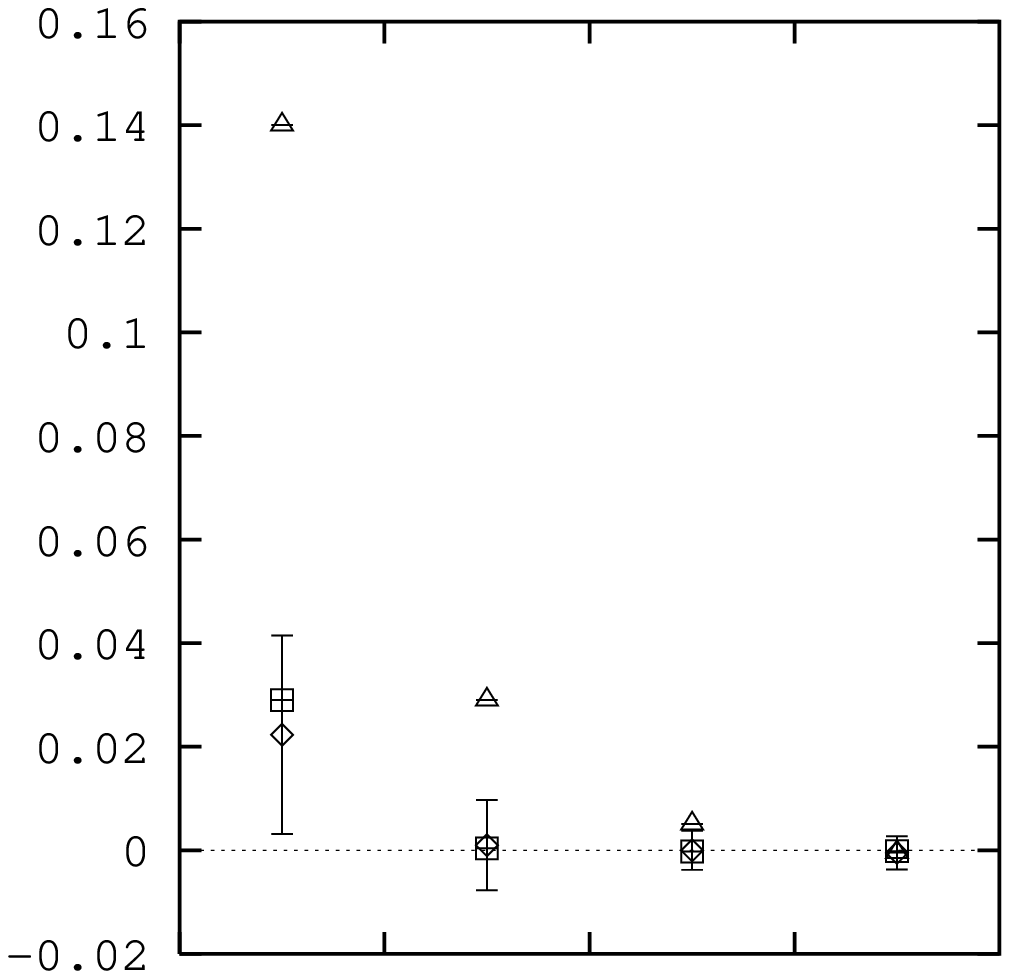,width=19cm,bbllx=2.cm,bblly=1.cm,bburx=21.cm,bbury=17.cm}}
 } \end{center}
\caption{b} 
\label{fig1b}
\end{figure}

 \newpage

\begin{figure}[p]
\begin{center} 
\mbox{\mbox{\epsfig{file=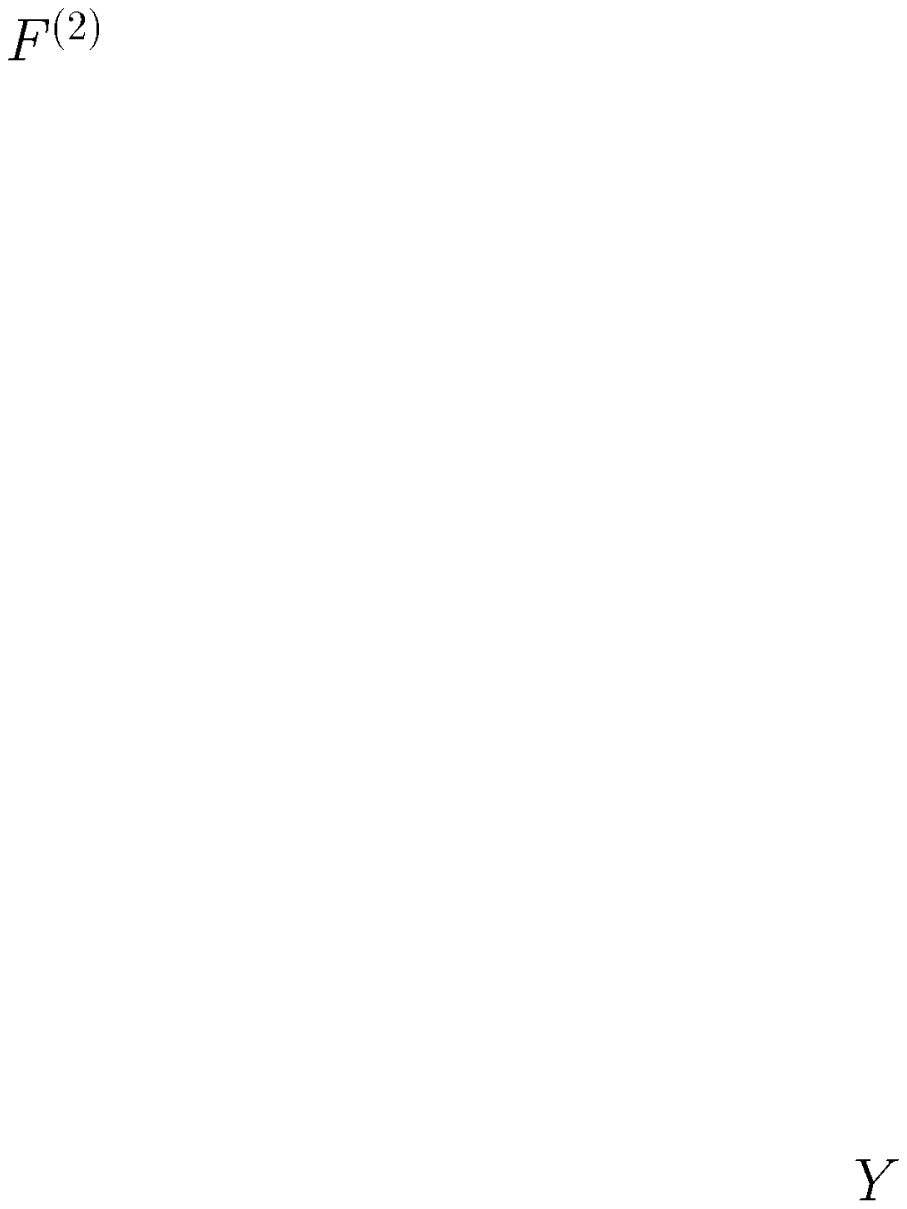,bbllx=4.cm,bblly=1.cm,bburx=4.2cm,bbury=17.cm}}
\mbox{\epsfig{file=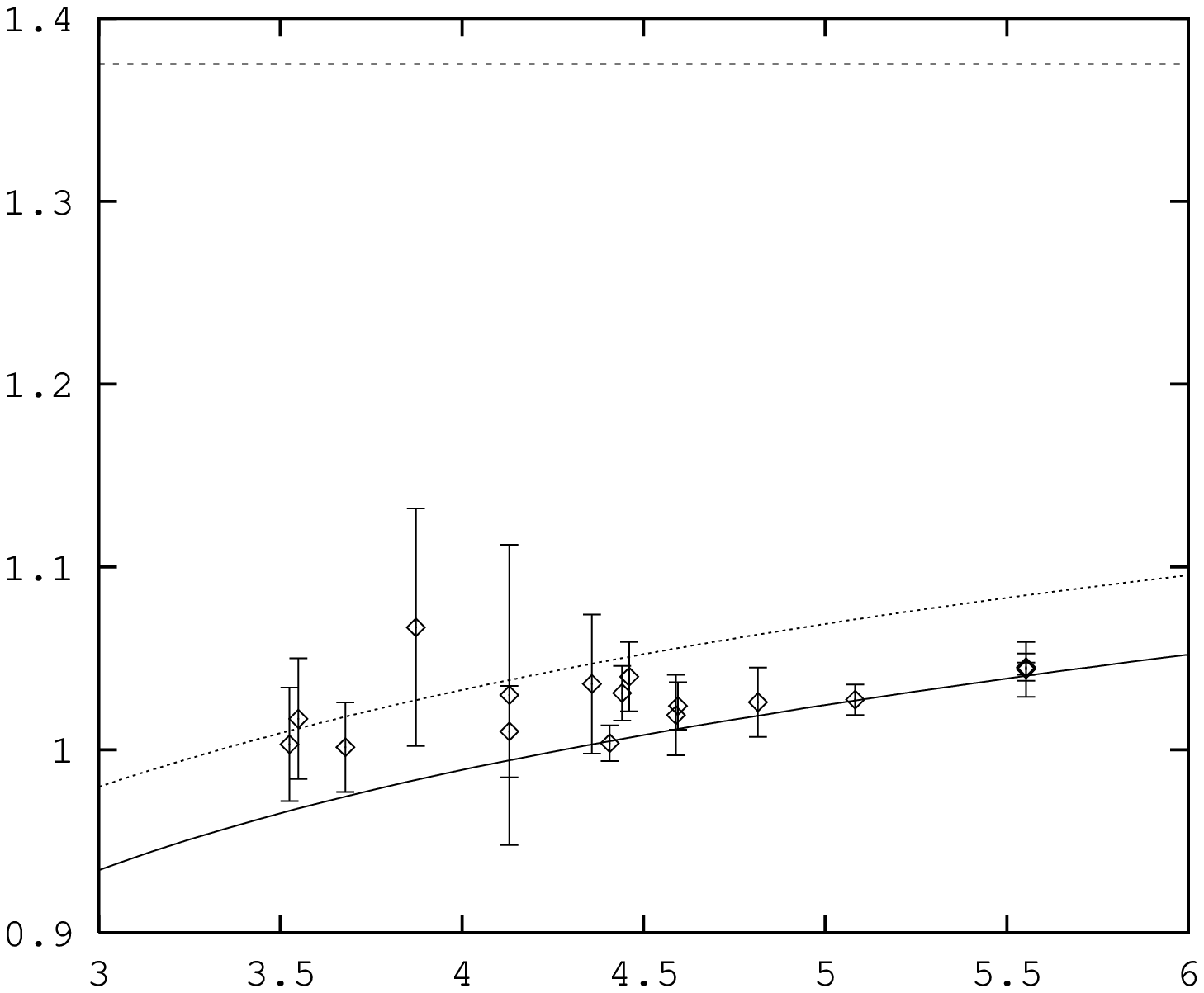,width=17cm,bbllx=4.cm,bblly=1.cm,bburx=21.cm,bbury=17.cm}}
 } \end{center}
\caption{} 
\label{figf2}
\end{figure}

 \newpage

\begin{figure}[p]           
\begin{center}
\mbox{\mbox{\epsfig{file=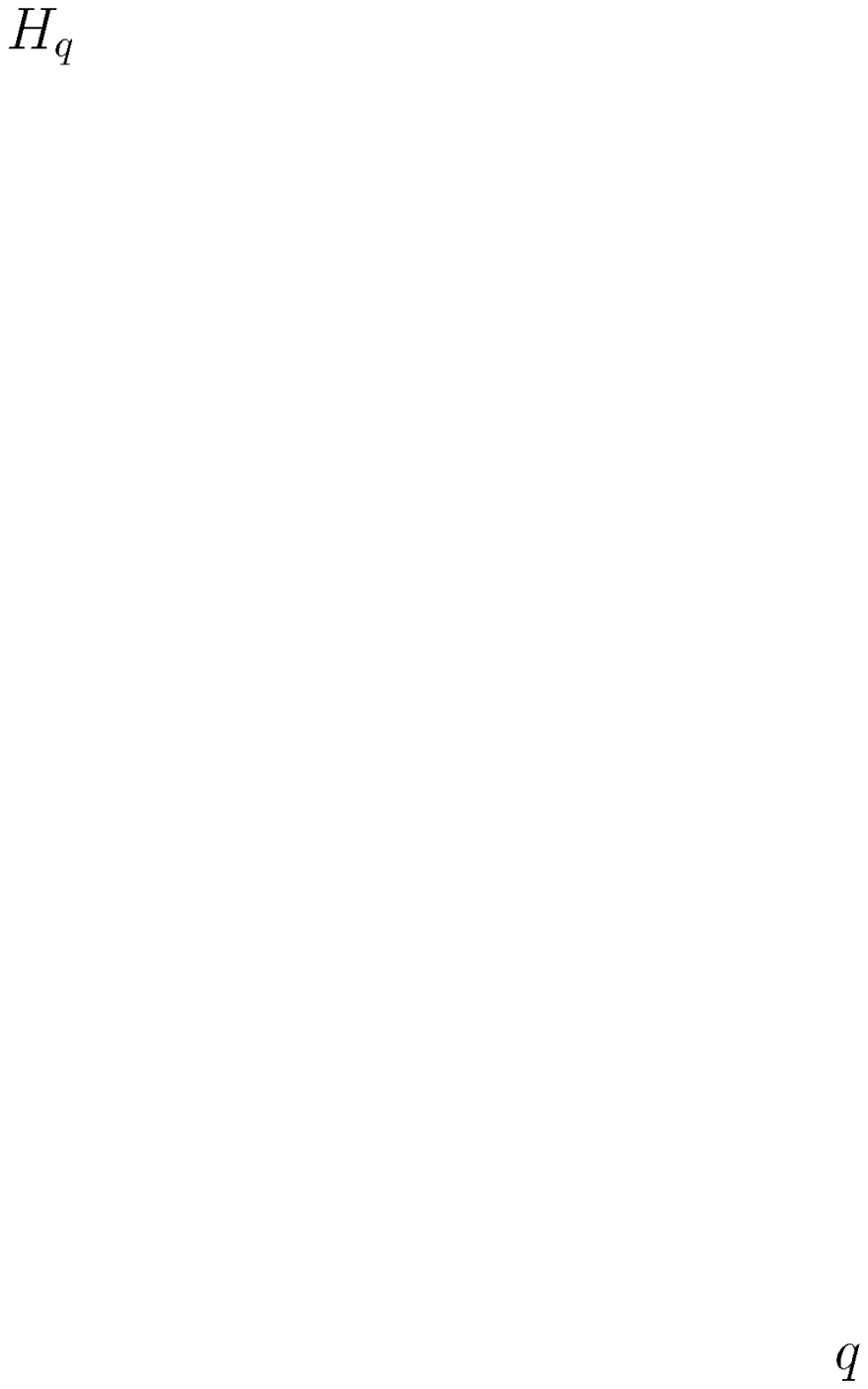,bbllx=4.cm,bblly=1.cm,bburx=4.2cm,bbury=17.cm}}
\mbox{\epsfig{file=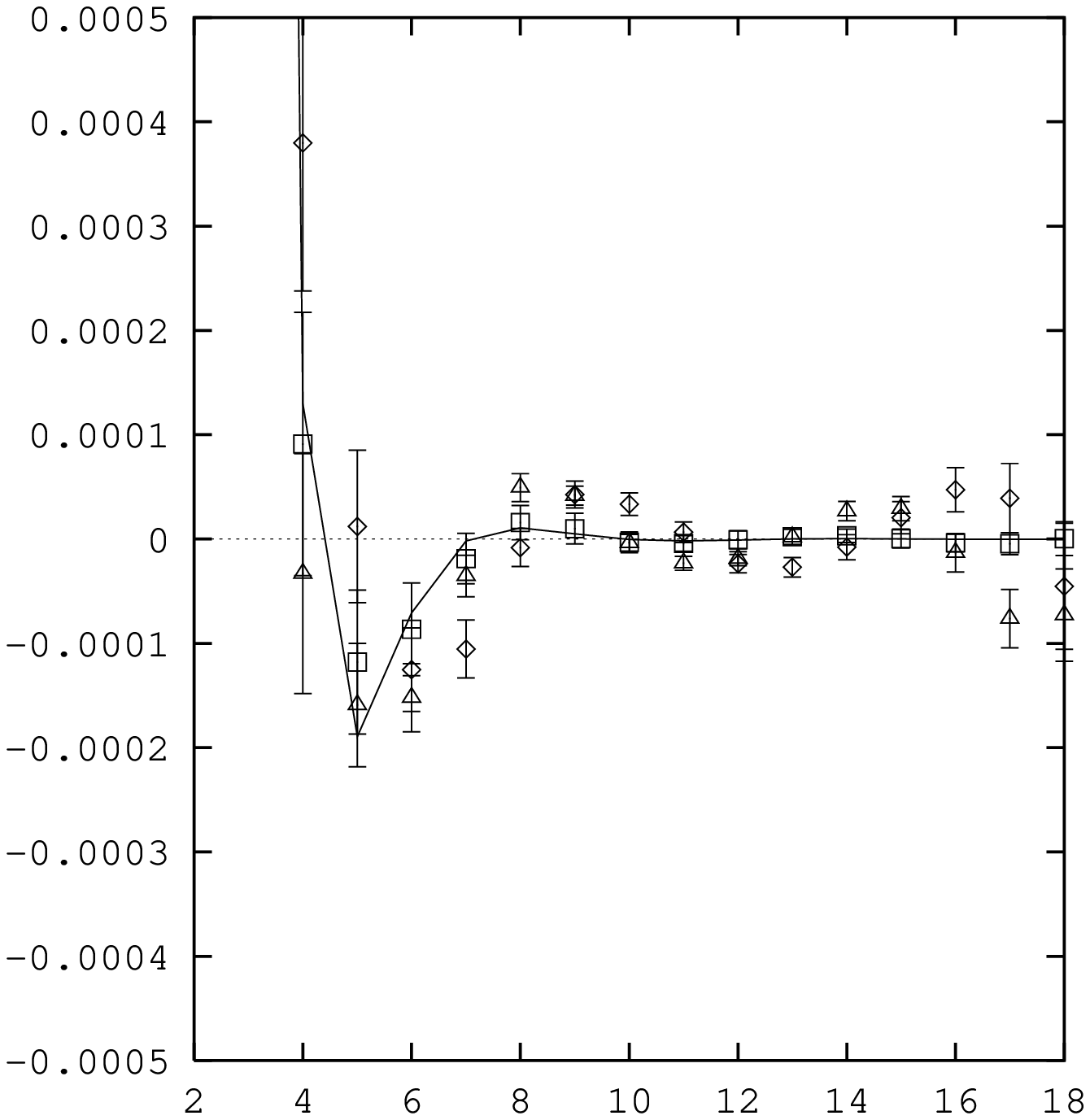,width=17cm,bbllx=4.cm,bblly=1.cm,bburx=21.cm,bbury=17.cm}}
 } \end{center}
\caption{}
\label{fighqduejet}
\end{figure}

\end{document}